\documentclass[11pt]{article}

\usepackage[letterpaper,margin=1in]{geometry}

\usepackage[T1]{fontenc}
\usepackage{lmodern}
\usepackage{microtype}
\usepackage{url}
\usepackage{graphicx}
\usepackage{booktabs}
\usepackage{pgfplots}
\pgfplotsset{compat=1.16}
\usepackage{array}
\usepackage{colortbl}

\definecolor{lkSD}{HTML}{B2182B}
\definecolor{lkD}{HTML}{EF8A62}
\definecolor{lkN}{HTML}{D9D9D9}
\definecolor{lkA}{HTML}{67A9CF}
\definecolor{lkSA}{HTML}{2166AC}
\definecolor{agree}{HTML}{4CAF50}
\definecolor{disagree}{HTML}{E8873A}
\newcolumntype{C}{>{\centering\arraybackslash}p{1.25cm}}

\newcommand\gencite[3]{\if\relax\detokenize{#3}\relax\cite{#1}\else\cite[#3]{#1}\fi}
\newcommand\genbibitem[3]{\bibitem{#1} #3}

\newcommand{\phasestablefloat}{%
\begin{table*}[t]
\centering
\caption{Phases of the ICSE 2026 Shadow PC.}
\label{phasestable}
\begin{tabular}{@{}p{0.18\linewidth}p{0.105\linewidth}p{0.635\linewidth}@{}}
\toprule
\textbf{Phase} & \textbf{Duration} & \textbf{Activities} \\
\midrule
HotCRP account setup & 1 week & Complete HotCRP profile including conflicts \\
Training and calibration & 3 weeks & Review 2 deliberately seeded papers on one of two topics; revise after checklist; synchronous Zoom discussion where everybody had reviewed the same two papers; breakout groups to critique reviews on those papers \\
Reviewing & 3 weeks & Bid and review 3 papers independently \\
Peer review of reviews & 1 week & Critique other reviewers' reviews \\
Final review & 2 weeks & Review 1 additional paper \\
Discussion & 2 weeks & Reach consensus; write meta-reviews \\
Debriefing / comparison & -- & Receive main PC reviews and author feedback for self-guided reflection \\
\bottomrule
\end{tabular}
\end{table*}%
}

\newcommand{\participantsurveyfloat}{%
\begin{figure*}[t]
\centering
\begin{tikzpicture}
\begin{axis}[
  xbar stacked,
  width=0.72\textwidth, height=5.4cm,
  bar width=11pt,
  xmin=-45, xmax=168,
  symbolic y coords={Q6,Q5,Q4,Q3,Q2,Q1},
  ytick=data,
  yticklabels={Learned about peer review process,
    Gained new perspective on PC work,
    Received useful feedback on reviewing,
    Found synchronous meeting useful,
    Found experience useful overall,
    Would recommend to others},
  y tick label style={font=\small},
  xtick={-40,-20,0,20,40,60,80,100},
  xticklabels={40,20,0,20,40,60,80,100},
  x tick label style={font=\footnotesize},
  xlabel={\small Percentage of respondents (diverging from Neutral)},
  extra x ticks={0}, extra x tick labels={},
  extra x tick style={grid=major, grid style={black!45, thin}},
  axis line style={draw=none},
  tick style={draw=none},
  enlarge y limits=0.14,
  clip=false,
  legend style={draw=none, at={(0.30,1.20)}, anchor=north,
    legend columns=5, column sep=5pt, font=\footnotesize},
  legend image code/.code={\draw[##1] (0cm,-0.07cm) rectangle (0.22cm,0.13cm);},
]
\addlegendimage{fill=lkSD,draw=black!45} \addlegendentry{Strongly disagree}
\addlegendimage{fill=lkD,draw=black!45} \addlegendentry{Disagree}
\addlegendimage{fill=lkN,draw=black!45} \addlegendentry{Neutral}
\addlegendimage{fill=lkA,draw=black!45} \addlegendentry{Agree}
\addlegendimage{fill=lkSA,draw=black!45} \addlegendentry{Strongly agree}
\addplot[forget plot,fill=lkN,draw=black!45] coordinates {(0,Q1)(-1.43,Q2)(-4.29,Q3)(-10.29,Q4)(-2.86,Q5)(-1.43,Q6)};
\addplot[forget plot,fill=lkD,draw=black!45] coordinates {(0,Q1)(-2.86,Q2)(0,Q3)(0,Q4)(0,Q5)(0,Q6)};
\addplot[forget plot,fill=lkSD,draw=black!45] coordinates {(0,Q1)(0,Q2)(0,Q3)(-2.94,Q4)(0,Q5)(0,Q6)};
\addplot[forget plot,fill=lkN,draw=black!45] coordinates {(0,Q1)(1.43,Q2)(4.29,Q3)(10.29,Q4)(2.86,Q5)(1.43,Q6)};
\addplot[forget plot,fill=lkA,draw=black!45] coordinates {(20,Q1)(31.43,Q2)(45.71,Q3)(32.35,Q4)(28.57,Q5)(22.86,Q6)};
\addplot[forget plot,fill=lkSA,draw=black!45] coordinates {(80,Q1)(62.86,Q2)(45.71,Q3)(44.12,Q4)(65.71,Q5)(74.29,Q6)};
\node[font=\small\bfseries] at (axis cs:126,Q1) [yshift=15pt] {Mean};
\node[font=\small\bfseries] at (axis cs:150,Q1) [yshift=15pt] {Agree};
\node[font=\small] at (axis cs:126,Q1) {4.80}; \node[font=\small] at (axis cs:150,Q1) {100\%};
\node[font=\small] at (axis cs:126,Q2) {4.54}; \node[font=\small] at (axis cs:150,Q2) {94\%};
\node[font=\small] at (axis cs:126,Q3) {4.37}; \node[font=\small] at (axis cs:150,Q3) {91\%};
\node[font=\small] at (axis cs:126,Q4) {4.15}; \node[font=\small] at (axis cs:150,Q4) {76\%};
\node[font=\small] at (axis cs:126,Q5) {4.60}; \node[font=\small] at (axis cs:150,Q5) {94\%};
\node[font=\small] at (axis cs:126,Q6) {4.71}; \node[font=\small] at (axis cs:150,Q6) {97\%};
\end{axis}
\end{tikzpicture}
\caption{Participant survey responses ($n=35$). The two right-hand columns give the mean (5\,=\,Strongly agree) and the fraction agreeing (``Agree'' or ``Strongly agree'').}
\label{participantsurvey}
\end{figure*}%
}

\newcommand{\icsecomparisonfloat}{%
\begin{table}[t]
\centering
\caption{Agreement between shadow PC recommendations and ICSE decisions. }
\label{icsecomparison}
\small
\setlength{\tabcolsep}{5pt}
\begin{tabular}{@{}l C C C @{\hspace{4pt}} r@{}}
\toprule
 & \multicolumn{3}{c}{\textbf{ICSE decision}} & \\
\cmidrule(lr){2-4}
\textbf{Shadow rec.} & \textbf{Accept} & \textbf{Major rev.} & \textbf{Reject} & \textbf{Total} \\
\midrule
Accept       & \cellcolor{agree!14}2 & \cellcolor{disagree!8}1  & \cellcolor{disagree!30}13 & 16 \\
Major rev.   & \cellcolor{disagree!8}1 & \cellcolor{agree!22}5  & \cellcolor{disagree!36}17 & 23 \\
Reject       & \cellcolor{disagree!20}6 & \cellcolor{disagree!20}6 & \cellcolor{agree!50}65 & 77 \\
\midrule
\textbf{Total} & 9 & 12 & 95 & 116 \\
\bottomrule
\end{tabular}
\end{table}%
}

\begin{document}
\title{The ICSE 2026 Shadow PC: Training the Next Generation of Reviewers Through Deliberate Practice}
\author{
Christian~Kästner, Michael~Hilton\\
Carnegie Mellon University
\and
Jonathan Bell\\
Northeastern University
\and
Francisco Gomes de Oliveira Neto\\
Chalmers University of Technology and\\the University of Gothenburg
}
\date{}

\maketitle

\begin{abstract}
        Peer review is essential to software engineering research, yet reviewer training remains largely implicit. We describe the ICSE 2026 Shadow PC, a redesigned program emphasizing \emph{deliberate practice} at scale. Key innovations include multi-phase structure with calibration and peer feedback, strict separation from the main PC, and a pathway toward leadership development. With 102 participants completing the program and reviewing 117 papers, 97\% recommending the experience, and positive reception from authors (67\% finding reviews helpful), the program demonstrates that rigorous reviewer training is achievable at scale. We share lessons learned and propose shadow PC area chairs as a mechanism for sustainable scaling and leadership development.
\end{abstract}

        \section{Introduction}\label{h.pfj8ydbvo013}
Peer review underpins scientific quality in software engineering, yet we have no systematic approach to training reviewers. Early-career researchers learn implicitly---by receiving reviews of varying quality and through occasional mentoring. Some advisors provide more direct instructions to their PhD students and enable various forms of subreviewing, but such opportunities are inconsistent and not available to all. Meanwhile, growing submission volumes demand more qualified reviewers and area chairs, but no deliberate pathway exists for developing them.

Shadow PCs have a long tradition as a professional development program for early-career scientists to develop their reviewing skills (and sometimes also to study the review process) \gencite{4DfUc,ssDk4,VUHpq}{[1, 2, 8]}{}. While there are several different formats, to the best of our knowledge, most shadow PC programs select a small number of participants, provide some upfront instructions through documents, videos, or remote meetings, and then have participants write reviews and discuss papers submitted to the conference. Sometimes the shadow PC organizers or more senior community members provide feedback on the reviews. In most cases, the shadow PC is strictly separated from the main PC and does not influence PC decisions, though some communities experiment with recruiting ``junior PC members'' directly into the main PC rather than having a separate one \gencite{hXq4M}{[7]}{}.

In the ICSE 2026 shadow PC, we aimed to overcome two limitations of many traditional shadow PC programs:

\begin{compactitem}
	\item Many shadow PCs limit the number of participants. For example, ICSE 2025's shadow PC selected only 50 of 299 applicants. We aimed to design a shadow PC that would provide professional development opportunities for all interested.

	\item Many shadow PCs provide some upfront instructions or guidelines and possibly some (often sparse) individual feedback later. While useful, we believe that we can design a better learning opportunity by grounding it in principles of \emph{deliberate practice}.

\end{compactitem}

We redesigned the ICSE 2026 Shadow PC around \emph{deliberate practice}: explicit training, iterative feedback, and reflection. Our contributions include:

\begin{compactitem}
	\item A multi-phase structure with calibration, peer review of reviews, and comparison with main PC reviews

	\item Execution at scale: 183 participants of which 102 participants completed the program, reviewing 117 papers, strict conflict-of-interest firewall from the main PC

	\item Evaluation via decision alignment, participant surveys (n=35), and author feedback (n=24)

	\item A path to continue to grow and expand this practice within our community

\end{compactitem}

\section{Program Design and Execution}\label{h.7hj14wp7girs}
\subsection{Guiding Principles}\label{h.4mnedy9uewp4}
Reviewing is fundamentally an exercise of \emph{evaluative judgment \gencite{sILRc}{[6]}{}}: the capacity to assess the quality of work against disciplinary standards and to articulate the reasons behind that assessment. Framing it this way suggests several approaches for cultivating judgment, on which we build: exemplars, explicit standards, calibration against expert assessment, and self- and peer-assessment.

These approaches share the logic of deliberate practice: repeated, goal-directed attempts with feedback and refinement, rather than one-shot performance. Concretely, we operationalized it through the following evidence-based teaching principles:

\begin{compactitem}
	\item \textbf{Generation before instruction.} Participants drafted reviews \emph{before} receiving checklists, exemplars, or guidance. Attempting the task first exposes the limits of the participants' current understanding, turning unconscious incompetence into conscious incompetence \gencite{lPsy8}{[4]}{ch.~5}. It makes the subsequent instruction more meaningful, the mechanism underlying "productive failure" \gencite{fl5jg}{[3]}{} and the case for a "just-in-time telling" \gencite{nFaNf}{[5]}{ch.~J}. We therefore made the first reviews low-stakes and expected them to be weak by design.

	\item \textbf{Contrasting cases and calibration.} To sharpen perception of what separates strong from weak papers, without being overly critical, a calibration phase contrasted a previously accepted paper and one with deliberately seeded flaws. In this phase, we had everyone review one of two  pairs of papers before we revealed expert judgments and discussed the divergence. Research argues that juxtaposing cases makes the dimensions of quality perceptible in a way that describing them in the abstract does not \gencite{nFaNf}{[5]}{ch.~C}.

	\item \textbf{Feedback paired with revision.} Rather than treat each review as a one-shot product, we design opportunities for feedback and revision. Research shows that feedback aids learning chiefly when it is timely and coupled with an opportunity to act on it \gencite{lPsy8}{[4]}{ch.~6} \gencite{nFaNf}{[5]}{ch.~F}.

	\item \textbf{Undoing misconceptions.} Novice reviewers arrive with predictable misconceptions or bad patterns, such as, enumerating issues without weighing their severity, looking for reasons to reject, or critiquing authors rather than the paper. We named these anti-patterns explicitly in instructions and through a checklist and had participants practice correcting them. This aligns with research on durable learning through confronting and revising a prior misconception rather than merely presenting the correct approach \gencite{nFaNf}{[5]}{ch.~U}.

	\item \textbf{Learning by teaching.} Critiquing others' reviews requires participants to articulate and apply the quality standards themselves, which reinforces their own evaluative judgment more than passively receiving those standards would. Research shows that the effort of teaching and evaluating others' work is itself a powerful driver of the teacher's own learning \gencite{nFaNf}{[5]}{ch.~T}.

	\item \textbf{Belonging and social learning.} Finally, a learner's sense of belonging, the feeling of being accepted and valued in a learning community, is a significant enabler of learning, independent of the subject matter. Research shows that when people feel they belong, they participate more fully, take intellectual risks, persist through difficulty, and engage more deeply \gencite{nFaNf}{[5]}{ch.~B} \gencite{lPsy8}{[4]}{ch.~7}. Fostering a sense of belonging among participants of the shadow PC was therefore a goal in its own right.

\end{compactitem}

In general, the program balanced realism with learning scaffolds: participants reviewed real ICSE submissions (from authors who opted in), but with explicit guidance, practice, checklists, and structured feedback unavailable in typical PC service. A strict conflict-of-interest firewall ensured shadow deliberations could not influence main PC decisions---shadow chairs and members had no access to main PC reviews until after notifications, and vice versa. We prioritized scale over exclusivity, accepting all qualified applicants rather than selecting an elite few.

\subsection{Program activities}\label{h.zaay1ld5jxqw}
We structured the shadow PC into multiple phases, summarized in Table~\ref{phasestable}, with deliverables at the end of each phase, such as completing their HotCRP profile, completing bidding, and completing the training reviews. At the end of each phase, we dropped all participants who have not completed the phase. In the end, 102 completed the entire program.

\phasestablefloat

\subsubsection{Phase 0: HotCRP Account Setup}\label{h.w35uzy9m5zj4}
In the first week, we asked all participants to complete their reviewer profile in HotCRP, including review preferences and conflicts. Mirroring reports from the PC chairs of the research track, many participants failed to fill out conflicts correctly (syntactically) and several participants required reminders or only completed the task after receiving an email about being dropped from the shadow PC. We noticed that this step alone is a useful forcing function and could benefit from more deliberate instruction.

\subsubsection{Phase 1a: Training and Calibration}\label{h.dxu4h0qnhilx}
In this phase, each shadow PC member reviewed two seeded papers (from a choice of two pairs). 

Our goal in designing the first phase was to use it as a formative assessment. Following the \emph{generation-before-instruction principle}, we first had the participants write a best effort version of their reviews without any further instructions. We did not expect these reviews to be of high quality, but they serve as a baseline for future improvement and allow for the participants to receive formative feedback.  After they completed their initial review, we then gave participants a checklist with guidance for good reviews (see Appendix) and asked them to refine their reviews, explaining what they did and did not change based on the checklist, providing them with an opportunity to receive feedback via their self-assessment. Logistically, we did this by asking participants to submit their initial reviews to a web form, which provided a link to the checklist and instructions for submitting the refined version to HotCRP, where they could see other reviews on the same paper.

For this step, we seeded papers following the \emph{contrasting-cases principle:} One paper in each pair was previously accepted to ICSE in the previous round,\footnote{ We asked a few authors whether they would volunteer their previously accepted paper for the shadow PC and would share the reviews they received. The papers were not formally published yet, but preprints may have been online. We asked the authors to change the paper title for the shadow PC version to make it less obvious.} and the other paper was a previously rejected submission in which we intentionally seeded additional problems.\footnote{ These two papers were provided and edited by the chairs of the shadow PC.} The intention of this selection was to provide some way of illustrating differences in reviews and calibrating expectations, where reviewers could compare their reviews with those of many others (we received 65+ reviews on each of these papers), where participants could see what kind of papers get accepted to ICSE some despite problems raised in reviews, and where we could lead a discussion of quite severe problems that the reviewers might have caught. This was intended to help with a common tendency of junior reviewers who might be overly negative and hold papers to unrealistic standards, or too shallow in their reviews. It enabled discussions about community standards for reviewing.

The calibration phase also had a side benefit regarding confidentiality: Starting with seeded papers allowed us to ensure that our participants are serious before they gain access to a large number of actual ICSE submissions. In fact, over 40 of our initial participants did not complete this phase.

\subsubsection{Phase 1b: Synchronous Zoom Meeting}\label{h.hjlkfgw3nhqp}
The synchronous zoom meeting served multiple purposes, from providing training materials, to calibrating expectations, to networking. We offered the meetings at two times to accommodate various time zones with two parallel tracks dedicated to the two pairs of seeded papers.

We started with a breakout where we asked participants in groups of five to introduce each other and then talk about the two papers they all reviewed. The purpose of this was to facilitate networking and belonging and to start seeing different arguments for the same work.

Afterward, we explained how we seeded papers and shared review fragments and seeded problems (emphasizing that they are intended for illustration of possible arguments from experts, but not intended as ground truth of how they should have reviewed the papers). This was intended to model behavior and calibrate expectations, especially to emphasize to not reject papers over minor problems when there are merits and to expect to recognize more severe issues.

We followed this with a presentation on the review process and some guidance on how to write high quality reviews.\footnote{ The slides we used are online at \url{https://docs.google.com/presentation/d/160_prBlEzxYZqTuLYu7hiaG1k45js-4IAtfuQ8eEZmI/edit?usp=sharing}}

To get participants to reflect about review quality and style, we asked them to contrast and critique three existing reviews with very different styles in a second breakout (with freshly shuffled groups to allow for more networking opportunities). To practice evaluative judgment, participants were asked to critique 3 reviews for one of the papers they have previously reviewed (contrasting cases again): One review was an original ICSE review, one review was a shallow AI generated review, and one review was an overly negative (hostile) review. 

We intended to have another session to apply the learning in improving a review together, but did not have sufficient time for it.

\subsubsection{Phase 2: Reviewing}\label{h.ig63qldjkqwi}
Next participants gained access to papers for bidding and were assigned three papers for reviewing. The intention of this phase was to practice the skills in writing reviews.

\subsubsection{Phase 3: Peer Reviews of Reviews}\label{h.irhpf71wz5jk}
Since providing individualized feedback by organizers or mentors was difficult to scale (over 350 reviews received), we decided to conduct peer reviews of the reviews, which allows participants both to practice critiquing reviews and also to receive feedback on their own. This design aligns with the learning-by-teaching principle. The peer feedback creates an external check and additional perspective on their writing, intended to enable additional reflection. We encouraged participants to use the checklist in their peer reviews. Each participant was asked to critique all reviews for two papers they had not reviewed, and each participant received feedback from two reviewers for each of their reviews. We asked participants to consider the feedback and revise their reviews where appropriate.

In our observation, most participants took the peer reviews seriously and provided constructive feedback.

\subsubsection{Phase 4: Final Review}\label{h.ayavpkjztd7b}
After previous phases of calibration and feedback, we asked participants to review one additional paper, intended to apply all their learning to write a high-quality review. At this point, we eliminated all scaffolding like peer review and trust participants to exercise their own learned judgment. 

\subsubsection{Phase 5: Discussion}\label{h.85cr4n54xxp2}
Finally, we asked participants to discuss the papers and reach consensus with a meta-review in a process that mirrored the actual ICSE process, with a discussion lead assigned for each paper. We provided some textual instructions on how to participate constructively in discussions and briefly covered this in our checklist (see appendix), but did not provide additional opportunities to practice.

\subsubsection{Phase 6: Debriefing / Comparison}\label{h.oattikhrudkf}
To overcome limitations from not being able to scale individualized feedback from organizers or mentors, we sought several ways to provide final feedback and closure. We received permission to share the anonymous reviews and meta-reviews from the ICSE PC on the same paper with the shadow PC reviewers so that they could compare and self-assess. We also invited the authors to provide feedback through HotCRP's rebuttal mechanism, though only few did.

All participants who completed all phases received a certificate of participation. We awarded 23 distinguished shadow PC reviewer awards, primarily based on high quality reviews and positive engagement during online discussions. We sent the list of distinguished reviewers to the PC chairs of ICSE 2027, ASE 2027, and FSE 2028, who invited many of them to their PCs.

Finally, we hosted a social session during an ICSE 2026 lunch break where we invited all shadow PC members that attended ICSE for some face-to-face interaction, furthering the belonging and social learning mission. Only a few attended, but it provided a nice opportunity for closure.

\subsection{Participants}\label{h.ofwpr2x68qps}
We opened the program to all who applied and meet the following criteria: \emph{PhD students, post-docs, new faculty members and industry practitioners working in software engineering research who (a) previously have received reviews for a paper submission in the technical research track (or the main track) of the premier SE conferences (e.g., ICSE, FSE, ASE), and (b) have not previously served as a program committee member of the technical research track of these conferences. [...] We particularly value participation from community members who may not have a strong local support network or institutional support for professional development (particularly review training) and are hence often overlooked in PC member selections.}

We intentionally tried to open participation to everybody interested, independent of local support, prior experience, or demographics. We did not ask for letters from advisors, evaluate motivation or other statements, or filter applications in other forms, but focused on designing a scalable experience instead.

We received 185 applications of which we invited all but two who failed to provide a valid email address in their application.

\textbf{Attrition:} Of 183 participants who started the program, 102 completed it. We dropped two participants for not declaring their conflicts of interest, 40 for not completing the first two reviews in the calibration phase, 11 for not bidding, 11 for not submitting phase 2 reviews, 17 for not completing the final review, and 7 for not participating in discussions. In a few cases, our email taking their lack of completing a step as a decision to withdraw triggered a response to complete the action late, on short notice, where we reinstated the participants. 

\subsection{Paper Sourcing}\label{h.xogwoxyeq7hn}
Authors of ICSE submissions could check a checkbox volunteering their paper for the shadow PC \emph{(``Shadow PC Opt-In: We confirm that we would like our paper to be considered for review in the Shadow PC track. ICSE 2026 features a Shadow PC track, i.e., a professional development program to train early-career researchers (PhD students, postdocs, new faculty members, and industry practitioners) in the review process of the technical track. More detailed information about the program is available at [...]. Shadow reviews for papers that are reviewed by the Shadow PC will be sent out to authors after the end of the actual review process; shadow reviews will not be considered in during the official review process made by the regular PC.'')}. Authors of 197 papers volunteered, of which 117 were selected based on conflicts and topic diversity. Papers creating conflicts with shadow PC chairs were excluded entirely.

\subsection{Evaluation}\label{h.lzwthapucskn}
We surveyed both participants and authors after the end of the shadow PC, receiving 35 responses from shadow PC members (34\% response rate) and 24 from authors (21\% response rate).

\subsubsection{Participant Survey (n=35)}\label{h.xd8z2ppdhvvb}
Shadow PC members (see Figure~\ref{participantsurvey}) were generally very positive about the program, indicating that they learned about the peer review process, gained a new perspective, received useful feedback, and found the overall experience useful. While not everybody found the synchronous meeting useful, it was often positively singled out in the feedback: \emph{"Having a face-to-face discussion on the merits of the papers was a really valuable experience."} Participants valued peer review of reviews (\emph{"an excellent learning opportunity"}) and the meta-reviewer role (\emph{"not common for other Junior PC tracks"}). Most found the 6-paper workload manageable, though some suggested reducing calibration to one paper. Concerns included unresponsive co-reviewers during discussions and desire for more connection to the actual PC.

\participantsurveyfloat

\subsubsection{Decision Alignment}\label{h.n1qge06u9a03}
We compared shadow recommendations against ICSE decisions for 117 papers, shown in Table~\ref{icsecomparison}. The shadow PC reliably identified weak submissions (68\% of ICSE rejects were also rejected by shadow) but was conservative on acceptance (only 2 of 16 shadow accepts matched ICSE). Shadow recommended major revision nearly twice as often as ICSE (23 vs. 12), possibly reflecting that on the main PC, this decision commits reviewers to re-review revisions, a cost shadow reviewers did not bear.

\icsecomparisonfloat

\subsubsection{Author Feedback (n=24)}\label{h.2u7uyrv18gmv}
Authors who volunteered papers were invited to provide optional feedback through the Rebuttal mechanism in HotCRP. Of 24 respondents, 67\% found reviews helpful, with a 5:1 ratio of positive to negative quality mentions. Strikingly, 63\% noted alignment between shadow and ICSE reviews, suggesting calibration succeeded not just in aggregate statistics but in specific critiques authors found salient. In addition, 58\% thanked reviewers, and 29\% said they would incorporate feedback. Author-reported decision alignment mirrored our quantitative analysis: The shadow PC was more conservative, rejecting 3 papers that ICSE accepted.

\subsection{Lessons Learned}\label{h.1ljiys1ve1wi}
We considered the redesign of the shadow PC program largely to be successful, though we also identified several opportunities for improvement.

\textbf{What worked:} We were able to scale the program to include all 183 applicants. The design to be inclusive but accept attrition worked. The initial calibration phase was effective at screening out participants who were not able or willing to invest time for the program. 

Multi-phase deliberate practice created visible learning progressions. The synchronous calibration meeting built community and was widely valued. Peer review of reviews distributed feedback burden while exposing participants to diverse approaches. 

Endorsement of distinguished reviewers in the shadow PC was appreciated by PC chairs of ICSE, FSE, and ASE and several participants have since been invited to the main PC of one or multiple of these conferences, providing a direct pathway to PC service, addressing the barrier of being unknown to those assembling committees. 

\textbf{Challenges:} Limiting the synchronous meeting to one hour left only limited time for deeper engagement with provided reviews and we had to skip the activity to collectively improve a review.

The main area of concern though was the discussion phase, where we provided some instruction but no opportunities for feedback and deliberate practice. Discussion quality varied widely---some papers had substantive debate while others stalled. Engagement decayed over the 3-month timeline, with discussion phases suffering most. Organizers wrote ~10 meta-reviews when discussions failed to converge or reviewers did not respond in discussions. Common issues included stating opinions as facts and meta-reviews that summarized rather than synthesized. During the discussion, four chairs managing over 100 reviewers was unsustainable, with the chairs having to intervene in many stalled discussions. No intermediate role existed between participant and chair for developing leadership skills.

The debriefing experience could have been improved. We posted the ICSE reviews and some (but not many) authors provided feedback but there was no subsequent engagement. The in-person session at ICSE 2026 was useful but only lightly attended.

\textbf{Recommendations:} Reflecting on our experience, we recommend the following for future Shadow PC offerings:

\begin{compactitem}
	\item \textbf{Invest in synchronous calibration:} The synchronous meeting is a high-value activity. Offer it for multiple time zones, record sessions, provide time for discussions among participants. It should not be shorter than one hour; longer sessions can provide more time for activities.

	\item \textbf{Improve discussion phase:} Apply design principles also to the discussion phase and provide opportunities for deliberate practice. On the instruction side, this should include setting explicit engagement expectations like expected response times in discussion. We also recommend conducting a second synchronous session to conduct a discussion and critique existing ones (contrasting cases) or model effective discussion behavior, followed by instructions and possibly another opportunity to practice. Another synchronous meeting might also re-energize participants.

	\item \textbf{Scale with area chairs:} Scale oversight of the discussion phase with area chairs. We propose shadow PC area chairs who each oversee 10-30 papers: monitoring discussions, nudging unresponsive reviewers, and ensuring meta-review quality. Recruitment pools include strong prior participants (creating advancement pathways) and junior main-PC members preparing for real AC duties (low-stakes practice). This creates a leadership pipeline: Shadow PC reviewer → Shadow PC Area Chair → Shadow PC Chair / Main PC Area Chair

	\item \textbf{Personalized discussion feedback:} If possible, provide personalized feedback on discussions via shadow area chairs or mentors

	\item \textbf{Debriefing:} Host a debriefing with exemplary examples and awards at ICSE.

	\item \textbf{Recognition:} Be explicit about recognition upfront. Identify awards criteria such as discussion quality, and explain pathways to PC service. Awards should be publicly shared.

	\item \textbf{Alumni community:} Shadow PC should be an entry point into a sustained community of practice. Alumni can serve as peer mentors, calibration facilitators, or area chairs---distributing burden while developing leadership skills. This way, each cohort can support the next.

	\item \textbf{Infrastructure:} HotCRP could better support shadow PCs through engagement dashboards, automated reminders, peer-review-of-reviews workflows, and hierarchical roles for area chairs.

	\item \textbf{Institutionalization:} Shadow PCs should become standard across ICSE, FSE, and ASE. Shared playbooks, reusable calibration materials, and cross-venue coordination would reduce startup costs and enable longitudinal tracking of alumni career trajectories.

\end{compactitem}

\subsection{Conclusion}\label{h.dao3eqjfo6p6}
The ICSE 2026 Shadow PC demonstrates that deliberate practice following evidence-backed teaching principles can transform reviewer training. With 102 participants completing the program, 97\% recommending the experience, reasonable decision calibration, and 67\% of authors finding reviews helpful, we consider the program successful in achieving both scale and quality. Challenges around engagement decay and organizer burden point toward shadow PC area chairs as a mechanism for sustainable scaling.

Beyond just reviewer training, the shadow PC has the potential to build leadership infrastructure. By creating tiered roles from participant to area chair, and publicly endorsing outstanding contributors for PC service, we establish a deliberate pathway for developing the judgment and organizational skills our community needs. As submission volumes grow, investing in this pipeline is essential to maintaining the integrity of peer review.

\subsection{Acknowledgments}\label{h.dim52tujzxa4}
We thank Tom Zimmermann and Mira Mezini, the ICSE 2026 PC Chairs for their partnership in navigating process complexities, facilitating paper volunteering, maintaining conflict-of-interest firewalls, and sharing main PC reviews. We also thank authors who volunteered papers and provided feedback, and all shadow PC participants.

\appendix

\section{Review Checklist (shared with participants)}\label{h.ab4eeo20rl97}
The following is a checklist for writing ICSE-style reviews. These are condensed from common recommendations by community members and shaped by the shadow-PC chair's own views.

Importantly, these are not hard requirements. You may intentionally violate them in individual cases and you may even fully disagree with some of them. 

\begin{compactitem}
	\item \textbf{1. Take a stance:} If you have expertise on the topic, try to take a clear stance for or against the paper. Expertise could be on the topic (e.g., static analysis) or the research method (e.g., qualitative analysis of interview data) or both. You are on the PC because others want to know your expert judgment. ``Weak accept'' and ``weak reject'' are more commonly used for reviews with less confidence, often outside your area of expertise, where you would like to hear what other reviewers say. Junior PC members may be experts on fewer papers, but they should still leverage their expertise where they are experts.

	\item \textbf{2. Clearly communicated reasons and priorities to authors and other reviewers:} Make sure your review makes it clear what arguments are important to you and how you reach your judgement. If you mention a number of positive and negative points, make clear how they balance. You can mention many minor points of criticism separately, but make sure that it is clear what your key concerns are (usually 2-3) if you argue against a paper. If you have criticism but argue for a paper overall, make sure you explain how the positives outweigh the negatives (e.g., frame the negatives as suggestions for improvements rather than serious problems). Authors of a paper may not agree with your conclusion, but it is vital that they understand your rationale. More importantly, other reviewers need to understand what your main concerns are so that they can focus on them in the discussion.

	\item It is a good strategy to add some narrative to a review. Clearly identify concerns and why they are severe. A mere list of unordered bullet points is an anti-pattern.

	\item \textbf{3. Summarize key points for easy skimming:} The ``positives'' and ``weaknesses'' fields are intended as \emph{redundant} summaries of your main points. They should be short bullet lists of a few keywords. They summarize what is explained in more detail in the main review; make sure everything mentioned in a bullet point is also covered clearly in the main review. The intention of these summary fields is to make it easy for other reviewers, area chairs, and PC chairs (and to a lesser degree also the authors) to skim reviews quickly and identify the overall arguments and shared themes or different positions among the reviews. Personally, I write these summaries at the very end to repeat the main points of your review, but you can also write them first and then use them as an outline for points to cover in your main review.

	\item \textbf{4. Cover all review criteria:} Many conferences explicitly indicate distinct review criteria; ICSE has four this year. Make sure all are covered in your review. You can but are not required to structure your reviews along those criteria. Personally, I prefer a review with a strong narrative focused on key points and tend to mention review criteria that did not strongly affect my judgement at the end for completeness. If you do structure your review by review criteria, ensure that it is possible to identify what is important and less important within each criteria and provide an overall summary judgement for the paper weighing the criteria at the end.

	\item \textbf{5. Try to be positive:} Almost every paper has some flaws, even award winning accepted ones. Try to review papers with a positive attitude and give papers a fair chance. Balance positives with negatives and deliberate about what problems are acceptable or can be easily fixed (e.g., presentation weaknesses, limited generalization, missing related work) and what problems are so serious that the paper should not be published in its current form (e.g., unsound claims, methodological flaws, missing novelty over prior paper). There is no limit on the number of papers that can be accepted to ICSE and no need to rank papers relative to other papers; instead find the line where papers can be accepted. Finding the line can require some practice, but try to err on the side of being positive. 

	\item Also remember that all study designs have tradeoffs and the authors may have deliberately focused on a different issue than what you would have liked (e.g., maximize control in an experiment with students over generalizability to industry practitioners). 

	\item Evaluate the paper how the authors designed and wrote it, not the paper you would have written.

	\item \textbf{6. Find the right length:} There is no right or wrong length of a review. Very negative reviews tend to be shorter: Find something positive to say, substantiate the main weakness, and feel free to skip minor comments. Positive reviews identify key contributions and their value and argue why weaknesses (if any) are acceptable, but may additionally have a long list of constructive suggestions for further improvements (often presentation related). Borderline papers often benefit the most from a reviewer's detailed, constructive feedback. When you write a long review, ensure that there is structure/narrative (of maybe 2-4 paragraphs) rather than just a long list of random issues. It is a good idea to separate minor points (e.g., typos, suggestions) from the core argument.

	\item \textbf{7. Be deliberate about questions:} ICSE allows reviewers to ask questions from the authors (not possible in the shadow PC). Only ask questions where the answer might change your judgment of the paper (e.g., clarifying an issue to see whether it is an easily fixable smaller writing problem). Be selective to allow authors to spend their time on important issues.

	\item \textbf{8. Critique the paper, not the authors:} Never talk about the authors in a review (e.g. ``The authors fail to explain'') but either judge the paper (e.g., ``The paper fails to explain'') or describe your own experience (e.g., ``I could not find/understand …'').

\end{compactitem}

Overall, develop your own style, but be deliberate about it.

\textbf{Additional guidelines for discussions}

\begin{compactitem}
	\item \textbf{A. Engage in discussions with substance:} Discuss with other reviewers about the strength and the weaknesses of the paper. Read the other reviews and respond where you agree or disagree with their key points, ask questions where answers might help move the discussion forward (by better understanding disagreements or reaching consensus). Engage with the other reviewers' arguments rather than just restating your own. Try to respond quickly (same day or next day) and constructively. 

	\item \textbf{B. Argue for your position:} You are on the PC because you have expertise. You have expertise and may see strengths and weaknesses that others with different backgrounds may not recognize. Even if the paper is not in your area of expertise, your opinion of how well the paper explains its contributions to an outsider is still valuable. If you think a paper makes a strong contribution and should be accepted (maybe despite some weaknesses), fight for it. If you think a paper is fundamentally flawed and should not be published in its current form, argue your position. Do not simply follow the majority or defer to more senior PC members.

	\item \textbf{C. Be open to changing your mind:} At the same time, listen to other reviewers and read the authors' response (if available). Be open to change your mind if you see other convincing arguments or find mistakes in yours. Maybe others see strengths that you did not appreciate. Explore tradeoffs and what weaknesses might be acceptable in a paper given other strengths, using all the information available.

	\item \textbf{D. Don't delay the inevitable:} If reviewers agree, there is no need to drag out the discussion. Summarize the key arguments across all reviews and see whether consensus can be reached quickly.

\end{compactitem}

\end{document}